\documentclass[letterpaper,10pt]{article} 

\usepackage{osameet3} 

\usepackage{amsmath,amssymb}
\usepackage[colorlinks=true,bookmarks=false,citecolor=blue,urlcolor=blue]{hyperref} 

\begin{document}

\title{Multi-color stimulated Raman scattering with a frame-to-frame wavelength-tunable fiber-based light source}

\author{Thomas Würthwein$^{1,*}$, Kristin Wallmeier$^{1}$, Maximilian Brinkmann$^{2}$, Tim Hellwig$^{2}$, Niklas M. Lüpken$^{1}$, Nick S. Lemberger$^{1}$, and Carsten Fallnich$^{1,3,4}$}
\address{$^1$ Institute of Applied Physics, University of Münster, Corrensstraße 2, 48149 Münster, Germany \\ 
$^2$ Refined Laser Systems GmbH, Mendelstraße 11, 48149 Münster, Germany \\
$^3$ MESA+ Institute for Nanotechnology, University of Twente, P.O. Box 217, Enschede 7500 AE, The Netherlands \\
$^4$ Cells in Motion Interfaculty Centre, University of Münster, Waldeyerstraße 15, 48149 Münster, Germany \\}
\email{*t.wuerthwein@uni-muenster.de}

\copyrightyear{2021}

\begin{abstract}
We present multi-color imaging by stimulated Raman scattering (SRS) enabled by an ultrafast fiber-based light source with integrated amplitude modulation and frame-to-frame wavelength tuning. With a relative intensity noise level of -153.7\,dBc/Hz at 20.25\,MHz the light source is well suited for SRS imaging and outperforms other fiber-based light source concepts for SRS imaging. The light source is tunable in under 5\,ms per arbitrary wavelength step between 700\,cm$^{-1}$ and 3200\,cm$^{-1}$, which allows for addressing Raman resonances from the fingerprint to the CH-stretch region.  Moreover, the compact and environmentally stable system is predestined for fast multi-color assessments of medical or rapidly evolving samples with high chemical specificity, paving the way for diagnostics and sensing outside of specialized laser laboratories.
\end{abstract}



\maketitle

\section{Introduction}

Stimulated Raman scattering (SRS) is advantageous for in vivo investigations or diagnostic imaging, due to the non-destructive, label-free and chemically selective nature of SRS and the accordingly low amount of required sample preprocessing. SRS has been applied successfully in cancer diagnostics \cite{Kendall2009, Yan2018, Huang2020}, stain-free histopathology \cite{Ji2013, Sarri2019, Li2019, Hollon2020} and other medically-relevant fields \cite{Ozeki2012, Suhalim2012, Shou2021a}. However, diagnostic SRS systems are rarely found in clinical settings, mainly due to the requirements concerning a high robustness and a small footprint of the required light source. 

For SRS imaging in laboratory environments typically mode-locked solid-state lasers in combination with synchronously pumped optical parametric oscillators (OPO) are used, as these allow for a synchronized dual-color output and a wide wavelength coverage. Furthermore, mode-locked solid-state lasers, i.e., titanium-sapphire lasers, benefit from a shot noise limited output at high frequencies \cite{Dietze2016, Audier2019}, making them in terms of noise performance ideal for SRS imaging with a high signal-to-noise ratio (SNR). However, the high complexity and demand for maintenance of the these light sources did not allow for a transfer to routine applications outside specialized laser laboratories.

In order to overcome the downsides of solid-state lasers in terms of the footprint and reliability for SRS imaging, fiber-based mode-locked light sources have been developed in the recent past~\cite{Nose2012a, Lamb2013, Freudiger2014, Kim2016, Ozeki2019, Kong2020}. However, these fiber-based light sources are typically based on a combination of ytterbium- and erbium-doped fibers, and, therefore, are limited in their tuning range to the CH-stretch region of the sample under investigation \cite{Freudiger2014, Kong2020, Audier2020}. Moreover, the noise of fiber-based mode-locked light sources is typically higher in comparison to solid-state lasers, which is most likely due to the high gain and the related strong amplified spontaneous emission in laser-active fibers \cite{Lamb2013, Kim2016}. The higher noise level in fiber-based light sources often deteriorates the SRS measurements, and, therefore, balanced detection has to be used for imaging with sufficient sensitivity \cite{Nose2012a, Freudiger2014}.

For sensitive SRS imaging the mode-locked light source presented in Ref.~\cite{Freudiger2014} showed a relative intensity noise (RIN) level of -140\,dBc/Hz and exploited balanced detection for the acquisition of SRS images at -167\,dBc/Hz, i.e. 3\,dB above the according shot noise. This setup was successfully applied for the differentiation between lipids and proteins in human tissue sections for virtual histology. Later on, a RIN of -147\,dBc at 20\,MHz was reached with an all-fiber laser system, consisting of mode-locked erbium- and ytterbium-doped fiber oscillators that were self-synchronized coherently \cite{Kong2020}. It was shown that this RIN level can be assumed as sufficiently low for imaging lipids in biological samples ranging from single cells to tissue sections without the need for balanced detection \cite{Kong2020}. However, aside from the achieved low RIN levels, promising for SRS microscopy of biological relevant samples, the application of the above listed fiber-based light sources suffered from a limited wavelength tunability within the CH-stretch region only, due to the limited gain bandwidth of the involved synchronized ytterbium- and erbium-doped fiber lasers. Moreover, the wavelength tuning was accomplished by a mechanically tunable slit and therefore was relatively slow, but might be optimized in a future setup~\cite{Kong2020}.

Here, we present high-speed multi-color SRS enabled by an ultrafast pulsed fiber-based light source, whose difference of its dual-color output is tunable between 700\,cm$^{-1}$ and 3200\,cm$^{-1}$, giving excess to the fingerprint region, the silent region for the investigation of deuterated samples~\cite{Zhang2020} and Raman tags~\cite{Li2017}, as well as the CH-stretch region. As for SRS imaging the noise performance of the light source is a crucial condition to measure the small modulation ($10^{-4} - 10^{-6}$, \cite{Rigneault2018,Audier2019}) that is transferred from the modulated pump wave to the unmodulated Stokes wave via the SRS process, we characterized the relative intensity noise (RIN) of the fiber-based light source. Moreover, multi-color SRS imaging, i.e., the acquisition of successive images at different wavenumbers, from the fingerprint to the CH-stretch region was performed with a tuning time of only 5\,ms for an arbitrary wavenumber step. The fast tuning of the light source allowed for frame-to-frame wavenumber switching at an imaging speed of 8\,Hz, limited only by the available non-resonant galvanometer scanner and without the need for balanced detection. Moreover, SRS spectra were acquired to prove the capability of the light source for SRS spectroscopy. The presented fiber-based light source, offering an alignment-free operation combined with a compact footprint, represents a promising tool to enable routine applications of multi-color coherent Raman imaging with a high chemical specificity for assessments of rapidly evolving (medical) samples.

\section{Fast and widely tunable fiber-based light source}

\begin{figure}[htbp]
\centering
\includegraphics[width=1\linewidth]{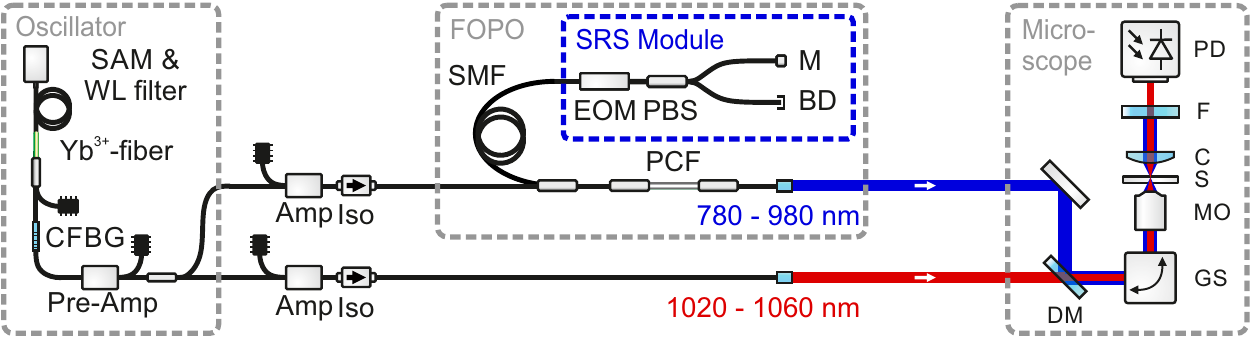}
\caption{Schematic of the dual-color output ultrafast fiber-based light source with integrated synchronized modulation at the half of the repetition rate of the fiber optical parametric oscillator (FOPO) for SRS microscopy with saturable absorber mirror (SAM), chirped fiber Bragg grating (CFBG), amplifiers (Amp), isolators (Iso), single-mode fiber (SMF), photonic crystal fiber (PCF), electro-optic modulator (EOM), polarizing beam splitter (PBS), mirror (M), beam dump (BD), galvo-scanner mirrors (GS), microscope objective (MO), sample (S), condensor (C), dichroic mirror (DM), filter (F), photomultiplier tube (PMT) and photodiode (PD).}
\label{fig1:laser}
\end{figure}

The presented light source was already investigated for coherent anti-Stokes Raman scattering (CARS) \cite{Brinkmann2019a} and frequency modulation CARS microscopy \cite{Wurthwein2021a}. The stated specifications hold for the here presented version of the light source, which was modified and optimized concerning a internal module for SRS. The fiber-based light source (Fig.~\ref{fig1:laser}) used an ytterbium-doped fiber oscillator as a seed source, which was mode-locked by a saturable absorber mirror and emitted pulses with a duration of 7\,ps and an output power of about 2\,mW. The repetition rate of the fiber oscillator was 40.5\,MHz at a central wavelength of 1040\,nm. A specifically chosen chirped fiber Bragg grating (CFBG) acted as the output coupler of the oscillator (transmission of 25\,\%) as well as a specific device to match the dispersion of the oscillator to the dispersion of the subsequent fiber-based optical parametric oscillator (FOPO) \cite{Brinkmann2016, Brinkmann2019a, Wurthwein2021a} to synchronize the repetition rates of both for different pump wavelengths. The central wavelength of the pulses could be tuned all-electronically between 1020\,nm and 1060\,nm in less than one millisecond by means of a custom-made fiber-coupled electronically tunable wavelength filter (switching time was verified by a measurement shown in Ref.~\cite{Brinkmann2019a}). 

After pre-amplification, the oscillator pulses were divided in power into two branches to seed two power-amplifiers based on double-clad ytterbium-doped fibers. The first branch was used for the application as Stokes pulses in SRS, which exhibited an average power of about 400\,mW, a duration of 7\,ps, and a spectral width of about 0.6\,nm. In the second branch the oscillator pulses were amplified to have sufficient energy (about 30\,nJ) to pump the subsequent FOPO. The linear cavity of the FOPO was formed by 50\,cm of polarization-maintaining (PM) photonic crystal fiber (PCF, NKT Photonics, LMA-PM-5) and about 155\,m of PM single-mode fiber (SMF, Nufern, PM780-HP) between a fiber-optical retroreflector and a polished FC/PC connector with a reflectivity of about 4\,\%. According to the resulting optical resonator length, the native repetition rate of the FOPO was 657\,kHz. Nevertheless, the repetition rate of the output signal pulses was solely determined by and equal to the repetition rate of the oscillator pulses of 40.5\,MHz, resulting in a harmonically pumped FOPO, which is analogous to the effect of harmonic mode-locking \cite{Zhou2006}. Thus, multiple signal pulses were circulating in the FOPO cavity with a temporal spacing equal to the temporal spacing of the oscillator pulses.

The particularly long fiber resonator was chosen to enable spectrally narrow dispersive tuning in the resonator \cite{Brinkmann2019a, Stolen1977, Yamashita2006}, such that the fed back signal pulses were temporarily stretched due to the group velocity dispersion of the fiber and only a narrow spectral part of the circulating signal pulses temporally overlapped with the subsequent oscillator pulses and were parametrically amplified. Due to this dispersive tuning, the full width at half-maximum (FWHM) bandwidth of the signal pulses was narrowed down to less than 12\,cm$^{-1}$ over the complete wavelength tuning range, which is well suited for highly selective SRS, matching the typical vibrational linewidths of biological samples in liquid environments. 

For the later SRS measurements an fiber-integrated electro-optic modulator (EOM) in combination with a  polarizing beam splitter (PBS) in the FOPO was used to modulate the amplitude of the circulating signal pulses at the half of the repetition rate, by sending every second pulse of the FOPO into the beam dump (BD) of the SRS module. The threshold behaviour of the FOPO was then exploited to achieve a 100\,\% modulation depth, which is typically not possible with external modulators \cite{Ozeki2010, Andresen2011, Steinle2019}, but beneficial for SRS imaging, as the amount of the SRS signal scales linearly with the modulation depth \cite{Audier2019}.

\section{Noise measurements}
For optimal SRS imaging results a low noise unmodulated beam is demanded and should be investigated in detail. The noise level of the modulated beam typically can be neglected \cite{Audier2019}, as it only transfers via the typically small Raman cross-section ($10^{-30}-10^{-28}$\,cm$^2$) to the SRS signal. Moreover, the modulation frequency of the modulated beam should be chosen to be in a frequency range with a low noise level. In order to quantify the noise performance of a light source typically the relative intensity noise \cite{Paschotta2004, Kim2016, Audier2019}  
\begin{equation}
    \text{RIN} = \frac{P_{\text{noise}}}{P_{\text{avg}}},
\end{equation}
with the noise power $P_{\text{noise}}$ and the average power $P_{\text{avg}}$ is determined. The RIN is ultimately limited at its lowest bound, i.e., $P_{\text{noise}} = P_{\text{shot}}$, given by the shot noise power 
\begin{equation} \label{equ:shotnoise}
 P_{\text{shot}} =2eIBR,
\end{equation}
where $e$ is the elementary charge, $I$ the photodiode current, $B$ the bandwidth of the measurement and $R$ the load resistance. For SRS imaging experiments the noise power should be as close as possible to the shot noise limit, as the SNR of the SRS signal is inversely proportional to the RIN \cite{Audier2019}.

\begin{figure}[htpb]
\centering
\includegraphics[width=1\linewidth]{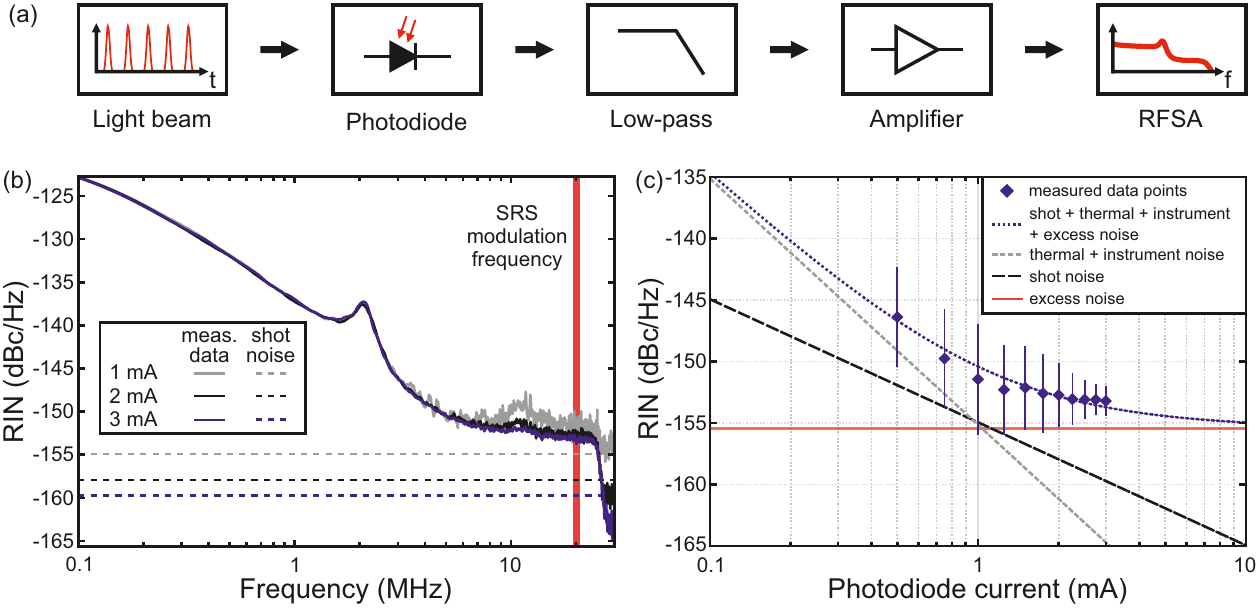}
\caption{(a) Schematic setup for the RIN measurements with a photodiode, low-pass filter (23\,MHz), a low-noise amplifier (20\,dB) and a radio-frequency spectrum analyzer (RFSA). (b) Frequency-resolved RIN of the Stokes output at 1039.5\,nm for 1\,mA, 2\,mA, and 3\,mA of photodiode current in comparison to the corresponding shot noise levels. (c) RIN for different photodiode currents of the Stokes output at 1039.5\,nm in comparison to the relevant noise contributions.}
\label{fig2:RIN}
\end{figure}

In order to quantify the applicability of the ultrafast fiber-based light source for SRS imaging, the RIN of this light source was measured as depicted in Fig.~\ref{fig2:RIN}(a), using a standard silicon photodiode (Thorlabs DET10A) with a load resistance of 50\,$\Omega$. The DC component, representing the average photocurrent, was measured with a multimeter for calculating with Eq.~\ref{equ:shotnoise} the corresponding shot noise limit. The AC component, representing the modulation on the DC signal, was filtered by a low-pass filter with a cut-off frequency at 23\,MHz, amplified by a home-built ultra-low noise amplifier (Texas Instruments, LMH6624, 20\,dB gain), and was characterized with a radio-frequency spectrum analyzer (RFSA, Anritsu MS2830A).

The RIN values are shown exemplarily for 1039.5\,nm wavelength (Fig.~\ref{fig2:RIN}(b)), which is in combination with the amplitude-modulated FOPO output the wavelength for addressing the vibrational resonance of dDMSO in the silent region for later SRS imaging experiments. The output at other wavelengths within the tuning range of the Stokes beam showed a similar RIN. The RIN values decreased up to 10\,MHz and stayed almost constant between 10\,MHz and the edge of the low-pass filter at 23\,MHz. The peak at 2\,MHz originated from relaxation oscillations of the oscillator~\cite{Weingarten1994}. With the frequency-resolved RIN measurements the modulation frequency for the later SRS measurements at 20.25\,MHz (red vertical line in Fig.~\ref{fig2:RIN}(b)) was confirmed to be within a suitable frequency band for low RIN values. 

The RIN values at the SRS modulation frequency of 20.25\,MHz are plotted for different photodiode currents in Fig.~\ref{fig2:RIN}(c) and showed a decrease for an increasing photodiode current, with a minimum of 153.7\,dBc/Hz at 3\,mA photodiode current. In addition, curves for the theoretical shot noise (black dashed line) and the estimated thermal noise \cite{Yoshimi2018,Ozeki2021} together with the measured noise of the instruments (grey dashed line) were added. Moreover, the blue dotted line was fitted to match the measured data points and represents the sum of all noise contributions. Two fit-parameters were used, from which the first describes the uncertainty of the thermal and instrument noise and the second the excess noise (red solid line), the latter being independent of the photodiode current. From this fit the excess noise could be determined to -155.5\,dBc/Hz. While excess noise originates from various sources as reviewed in Ref.~\cite{Kim2016}, it seems that the low noise level can be attributed to the narrowband spectral filtering in the laser oscillator, as we found that an increase of the oscillator bandwidth from about 0.1\,nm to 0.3\,nm led to an increased excess noise level from 155.5\,dBc/Hz to -152.5\,dBc/Hz. A similar behavior was observed in other mode-locked lasers~\cite{Qin2014, Shin2015} and described theoretically~\cite{Haus1993,Paschotta2004}. As the excess noise limits the RIN of the unmodulated Stokes beam, a further increase of the photodiode current would not increase the image quality in the later SRS images \cite{Audier2019}. 

The RIN level of the amplitude-modulated FOPO output (pump beam in SRS terminology) was at about -126\,dBc/Hz, i.e., almost 28\,dB above the noise level of the Stokes beam. This increase of the RIN was also observed in mechanically tunable FOPO \cite{Lamb2013}, and is most likely attributed to the high FWM gain combined with the low feedback efficiency (approx. 4\,\%) of the FOPO resonator. However, the RIN of the unmodulated beam has the highest priority and was with 153.7\,dBc/Hz on a suitable level for SRS imaging. For comparison, SRS imaging of tissue sections was successfully shown using a light source with a 6\,dB higher noise level of the unmodulated beam \cite{Kong2020}. Following this argumentation, the presented fiber-based light source should enable sensitive SRS imaging by detecting the stimulated Raman gain on the Stokes beam.


\section{SRS imaging and spectroscopy}

\begin{figure}[htpb]
\centering
\includegraphics[width=1\linewidth]{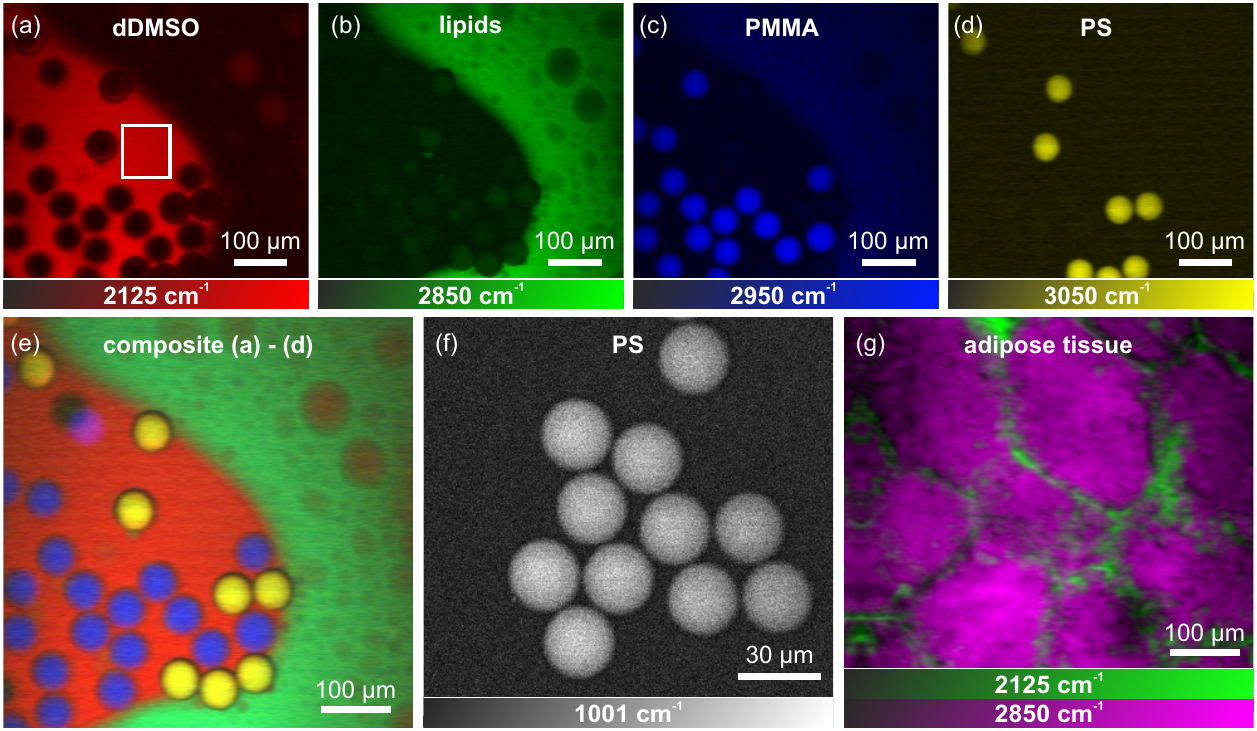}
\caption{SRS images of (a) deuterated dimethyl sulfoxide droplets at 2125\,cm$^{-1}$ (dDMSO, red), (b) lipids at 2850\,cm$^{-1}$ (green), (c) poly(methyl methacrylate) at 2950\,cm$^{-1}$ (PMMA, blue), (d) polystyrene at 3050\,cm$^{-1}$ (PS, yellow), and (f) PS at 1001\,cm$^{-1}$ (white). (e) The multi-color composite represents the superimposed images of the components (a) to (d) and the marked image area in (a) was used to determine the image quality. (g) Dual-color SRS images of adipose tissue soaked with dDMSO. All images (1024x1024 pixels, five averages) were acquired with pixel dwell times of 10\,µs.}
\label{fig3:SRS_Images}
\end{figure}

In proof-of-principle multi-color imaging experiments, the output of the FOPO was used as pump beam and the amplified pulses from the oscillator as Stokes beam for SRS microscopy. Pump and Stokes pulses, overlapping in time and in space at the dichroic mirror (DM), were coupled into a home-built laser scanning microscope (see Fig.~\ref{fig1:laser}), consisting of a pair of non-resonant galvo-scanners (GS) and a 60x, 0.75\,NA microscope objective (MO). The beams, transmitted by the sample (S), were collected with a condenser (C, NA $=0.79$), and the Stokes beam was spectrally filtered out by two optical long-pass filters (F, long-pass 900\,nm) and relayed onto a home-built detector (PD) for Stokes signal acquisition. The home-built detector was designed along the recommendations from Ref.~\cite{Gray1998}, while a large-area photodiode (Hamamatsu, S3590-08) and a ultra-low noise wideband amplifier (Texas Instruments, LMH6624) were used. The AC signal of the detector was low-pass filtered (cut-off at 23\,MHz) and analyzed with a lock-in amplifier (Zurich Instruments, HF2). 
The average output powers were adjusted to stay well below the damage threshold of the samples and were adjusted to maximize the SRS signal to roughly 32\,mW and 46\,mW for the pump and Stokes beams, respectively. This Stokes beam power corresponded to 15\,mW in front of the detector, resulting in 3\,mA of photodiode current. The incident power on the sample can be further reduced by reducing the losses in the detection path or by using a photodiode with a higher responsivity. The multi-color SRS images in Fig.~\ref{fig3:SRS_Images} (1024x1024 pixels with a pixel dwell time of 10\,µs) were acquired by frame-to-frame wavelength switching, i.e., changing the excitation wavenumber between consecutive images; and averages were calculated via the arithmetic mean of the complete images. The settings of the light source and the image acquisition were controlled by a self-written MatLab program.

Switching the excitation wavenumber allowed the differentiation of (a) deuterated dimethyl sulfoxide droplets at 2125\,cm$^{-1}$ (dDMSO, red), (b) lipids at 2850\,cm$^{-1}$ (green), (c) poly(methyl methacrylate) at 2950\,cm$^{-1}$ (PMMA, blue), and (d) polystyrene at 3050\,cm$^{-1}$ (PS, yellow). The composite of the acquired images (a) to (d), shown in Fig.~\ref{fig3:SRS_Images}(e), demonstrates the chemical specificity enabled by the presented light source.
The image quality was quantified by the $SNR =  \langle S\rangle / \sigma(S)$, with $\langle S \rangle$ being the average and $\sigma(S)$ the standard deviation of a signal $S$, measured in an image area with a homogeneous signal of dDMSO (white rectangle in Fig.~\ref{fig3:SRS_Images}(a)). In the labeled image area the SNR was between 14.2 (no averaging) and 29.2 (five averages), and thus in a comparable magnitude to other SRS experiments \cite{Heuke2020}.
In order to further test the applicability of the ultrafast fiber-based light source, PS in the fingerprint region at 1001\,cm$^{-1}$ (Fig.~\ref{fig3:SRS_Images}(f)), and adipose tissue (soaked with dDMSO for two hours) was imaged (Fig.~\ref{fig3:SRS_Images}(g)), addressing the molecular vibrations of dDMSO at 2125\,cm$^{-1}$ (green) and lipids at 2850\,cm$^{-1}$ (magenta). The pixel dwell time was 10\,µs and the images were averaged five times to improve the SNR. 
Furthermore, to demonstrate the ability of the light source to study moving samples with frame-to-frame wavelength switching, we have attached a video (\textcolor{blue}{Visualization~1}) to the supplementary material showing real-time imaging at 8\,fps (limited by the used linear galvo scanners) and without signal averaging of a moving dDMSO/oil sample. In particular, this measurement emphasizes the benefits of the here presented light source, as it was possible to switch between arbitrary Raman resonances, here the dDMSO resonance at 2125\,cm$^{-1}$ and the oil resonance at 2850\,cm$^{-1}$, in less than 5\,ms. The fast wavelength tunability of the light source makes it ideal for the rapid assessment and analysis of moving or fast evolving samples.

\begin{figure}[htpb]
\centering
\includegraphics[width=1\linewidth]{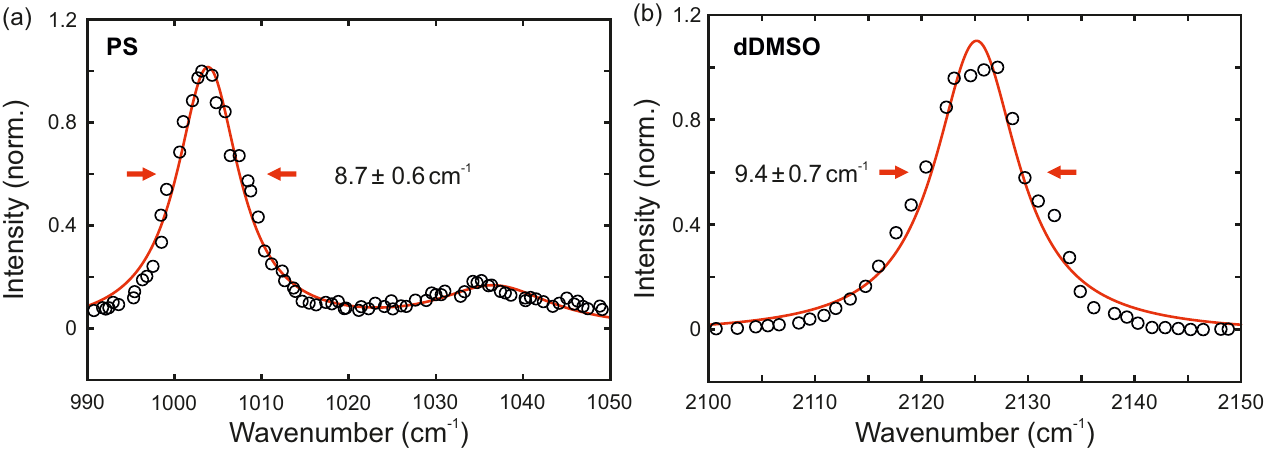}
\caption{SRS spectra of (a) PS beads in the fingerprint region between 990\,cm$^{-1}$ and 1050\,cm$^{-1}$ and (b)~dDMSO around 2125\,cm$^{-1}$. The resonances showed a spectral width of 8.7\,cm$^{-1} \pm$\,0.6\,cm$^{-1}$ and 9.4\,cm$^{-1} \pm$\,0.7\,cm$^{-1}$, respectively.}
\label{fig4:SRS_spectra}
\end{figure}

In order to prove the capability of the light source for the acquisition of SRS spectra, the resonances of PS between 990\,cm$^{-1}$ and 1050\,cm$^{-1}$ and the resonance of dDMSO at 2125\,cm$^{-1}$ were measured (Fig.~\ref{fig4:SRS_spectra}(a) and (b)). High-resolution SRS spectra were acquired with a precise wavenumber tuning around the resonances, which was accomplished by keeping the oscillator wavelength fixed while adjusting the length of the FOPO via the position of mirror M. Spectra were extracted from images at different wavenumbers, by integrating over the background-corrected intensity data and by normalizing them to the incident laser power at the according wavenumber. The resulting spectra follow a Lorentzian lineshape with a FWHM of 8.7\,cm$^{-1} \pm$\,0.6\,cm$^{-1}$ and 9.4\,cm$^{-1} \pm$\,0.7\,cm$^{-1}$, respectively, which is in compliance with literature~\cite{Vibrational1981, Martens2002}. The resonance of PS at 1035\,cm$^{-1}$ was hardly resolvable and showed a FWHM of  16.3\,cm$^{-1}$ with a relatively large error of $\pm$ 6.2\,cm$^{-1}$.

\section{Conclusion}
In conclusion, a rapidly and widely tunable ultrafast light source for multi-color stimulated Raman scattering (SRS) was presented. A mode-locked fiber oscillator with a subsequent fiber optical parametric oscillator (FOPO) provided the tuning capability from the fingerprint to the CH-stretch region (700\,cm$^{-1}$ to 3200\,cm$^{-1}$). The rapid tunability in 5\,ms for an arbitrary wavenumber step enabled live multi-color imaging with SRS at a frame rate of 8\,Hz, changing between CD- and CH-stretch vibrations in a frame-by-frame manner. Moreover, the capability of the light source for the acquisition of SRS spectra was proven by measuring resonances of PS and dDMSO in the fingerprint and silent region, respectively. The noise properties of this fiber-based light source were shown to be favorable for SRS imaging and spectroscopy without the need for balanced detection with a low relative intensity noise level of the unmodulated Stokes output beam at about -153.7\,dBc/Hz. Beside high-speed and wide tuning as well as low-noise performance the light source shows a compact footprint and a robust operation, making it suitable for future applications in challenging environments, e.g. medical practices, clinics, and operating rooms.

\section*{Acknowledgments}
The authors gratefully thank Ingo Diedrich for his strong support in the fabrication of the detector used for the SRS measurements.

\section*{Disclosures}
 MB and TH: Refined Laser Systems GmbH (I,E)

\section*{Supplemental document}
See Visualization 1 for supporting content.

\end{document}